\documentclass[osajnl,twocolumn,showpacs,superscriptaddress,10pt]{revtex4-1} %% use 11pt for Applied Optics
\usepackage{amsmath,amssymb,graphicx}

\usepackage{subcaption}
\begin{document}

\renewcommand\arraystretch{0}
\renewcommand\tabcolsep{0pt}

%\title{Ultra-low-loss Waveguide Crossing Arrays Based\\on Imaginary Coupling of Multimode Bloch Waves\vspace{-10pt}}
%\title{Ultra-low-loss Waveguide Crossing Arrays Based\\on Multimode Bloch Waves and Imaginary Coupling\vspace{-10pt}}
\title{Ultra-low-loss CMOS-Compatible Waveguide Crossing Arrays\\Based on Multimode Bloch Waves and Imaginary Coupling\vspace{-10pt}}

\author{Yangyang Liu, Jeffrey M. Shainline, Xiaoge Zeng and Milo\v{s} A. Popovi\'{c}}
\affiliation{Department of Electrical, Computer, and Energy Engineering, University of Colorado, Boulder, CO  80309, USA
\vspace{-5pt}}

%\author{Yangyang Liu}
%\affiliation{Department of Electrical, Computer, and Energy Engineering, University of Colorado, Boulder, CO  80309, USA}
%
%\author{Jeffrey M.  Shainline}
%\affiliation{Department of Electrical, Computer, and Energy Engineering, University of Colorado, Boulder, CO  80309, USA}
%\affiliation{Quantum Electronics and Photonics Division, National Institute of Standards and Technology, Boulder, CO 80305, USA\\
%$^*$Email: milos.popovic@colorado.edu
%\vspace{-5pt}}
%
%\author{Milo\v{s} A. Popovi\'{c}$\,^{1,*}$}
%%\affiliation{Department of Electrical, Computer, and Energy Engineering, University of Colorado, Boulder, CO  80309, USA}

\begin{abstract}
\vspace{-30pt}
We experimentally demonstrate broadband waveguide crossing arrays showing ultra low loss down to $0.04\,$dB/crossing ($0.9\%$), matching theory, and crosstalk suppression over $35\,$dB, in a CMOS-compatible geometry.  The principle of operation is the tailored excitation of a low-loss spatial Bloch wave formed by matching the periodicity of the crossing array to the difference in propagation constants of the 1$^\text{st}$- and 3$^\text{rd}$-order TE-like modes of a multimode silicon waveguide.  Radiative scattering at the crossing points acts like a periodic imaginary-permittivity perturbation that couples two supermodes, which results in imaginary (radiative) propagation-constant splitting and gives rise to a low-loss, unidirectional breathing Bloch wave.  This type of crossing array provides a robust implementation of a key component enabling dense photonic integration.% in a single lithographic step.  The underlying Bloch wave concept suggests a path to higher-order, more efficient designs, rigorous design methods, and applications to thermal tuning, modulators and optomechanics.
%, and is stabilized by potential scattering at the junctions, which can be understood as an imaginary-valued coupling mechanism leading to imaginary splitting of propagation constants.  The periodicity of the crossings is designed to match 
\end{abstract}

\maketitle %% required

%%%%%%%%%%%%%%%%%%%%%%%%%%%%%%%%%%%%%%%%%%%%%%%%

%\section{Introduction}

\noindent Silicon photonics is beginning to enable complex on-chip optical networks comprising hundreds of devices.  One emerging application is energy efficient, chip-scale photonic interconnects for CPU-to-memory communication \cite{BattenIEEEMicro2009}.  With increasing device density and complexity in a planar photonic circuit, efficient waveguide crossings are indispensible in many network topologies %For some network topologies, the number of waveguide crossings required rises quickly with circuit size and tolerable levels of loss and crosstalk per crossing accordingly drop to very small limits
 \cite{BattenIEEEMicro2009}.  %A multitude of work has considered crossing designs over the past decade. %\cite{Manolatou:99,Stuart:03,Liu:04,Fukazawa:05,Bogaerts:06,awpoon,Barwicz:07,Popovic07xing,Hochberg}, including resonant \cite{Manolatou:99}, MMI-like \cite{Stuart:03,Liu:04,awpoon,Popovic07xing,Hochberg}, multi-layer\cite{Bogaerts:06}, and adiabatic \cite{Fukazawa:05,Barwicz:07} designs.
%Crossing designs based on adiabatic tapers have demonstrated modest insertion loss, in the range of 0.3--1\,dB, and have a large device footprint \cite{Fukazawa:05,Barwicz:07}.  Resonant designs permit low loss and crosstalk in a compact footprint, but have narrow bandwidth \cite{Manolatou:99} (e.g. $\sim\!4\,$nm \cite{Yu:12}).  Multilayer processes allow crossing waveguides with reduced scattering \cite{Bogaerts:06}, or complete separation through vertical displacement in separate material layers \cite{Jones:13}.  These approaches bring with them the complexity of multiple lithographic steps and material layers, and in the latter case may require interlayer couplers with additional loss and footprint penalties.  
Crossing designs based on adiabatic aperture widening are large and relatively lossy (0.3--1\,dB) \cite{Fukazawa:05,Barwicz:07,WattsXing}, while resonant designs permit low loss and crosstalk in a compact footprint, but have narrow bandwidth \cite{Manolatou:99} (e.g. $\sim\!4\,$nm \cite{Yu:12}).   
Multilayer processes allow reduced scattering in crossing waveguides \cite{Bogaerts:06} or their complete isolation through vertical displacement \cite{Jones:13}, but they require multiple lithographic steps and/or material layers.  %, and in the latter case may require interlayer couplers with additional loss and footprint penalties.  
Multimode-interference (MMI) based crossings \cite{Stuart:03,Liu:04,Poon:06,Popovic07xing,Popovicxingpatent,Hochberg}, despite ostensibly multimode behavior, have a number of attractive features, with individual crossings down to $0.18\,$dB loss and $41$\,dB crosstalk \cite{Hochberg}.

\begin{figure}[b]
\vspace{-8pt}
\centering
\includegraphics[width=\linewidth]{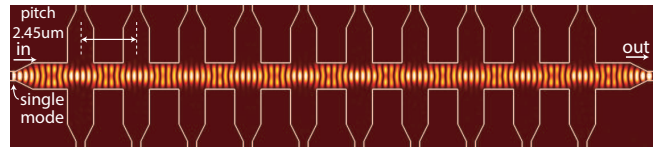}
%\label{good}
%\label{bad}
\caption{Ultra-low-loss waveguide crossing array, based on excitation of a low-loss breathing Bloch wave, formed of 1$^\mathrm{st}$ and 3$^\mathrm{rd}$ modes of a multimode waveguide (2D FDTD simulation) \cite{Popovic07xing}.  The mode is stabilized by radiative loss.\label{withtaper}}
\vspace{-6pt}
\end{figure}

In this Letter, we describe ultra-low-loss waveguide crossing arrays based on a periodic multimode structure.  Popovi\'{c} \emph{et al.} \cite{Popovic07xing} proposed an efficient approach to design a crossing array (Fig.~\ref{withtaper}) by constructing a low-loss Bloch wave in a matched periodic structure where the optical field synthesizes periodic focii that jump across gaps and avoid diffraction loss and scattering at the crossing points.
%, and where non-adiabatic tapered excitation structures are used to efficiently excite these low-loss Bloch waves
 This concept is reminiscent of periodic lens-array microwave beam guiding \cite{Goubau:61}.  Microphotonic implementations use a minimum of modes to implement focusing physics, eliminate reflections, and introduce new degrees of freedom.  In the first experimental demonstration of this concept \cite{Liu:13}, we showed record low waveguide-crossing loss of $0.04$\,dB/crossing (0.9\%), equal to theoretical design efficiency \cite{Popovic07xing}. %, demonstrating the robustness of these structures.
Another recent paper \cite{Zhang:13} demonstrated similar crossing arrays based on our proposal in Ref.~\onlinecite{Popovic07xing}, achieving 0.14\,dB loss, and introduced an improvement based on subwavelength patterning of the sidewalls, reducing the loss further to below $0.02$\,dB.  To our knowledge, these two results represent respectively the lowest achieved crossing loss in CMOS-compatible photolithography processes \cite{Liu:13} and in high-resolution processes allowing nanopatterning, such as scanning electron-beam lithography \cite{Zhang:13}.  
In this paper, we present the low-loss unidirectional Bloch mode concept, its use in design of waveguide crossing arrays, and our experimental results on record loss in CMOS-compatible crossings.  We show an excellent match between theory and experiment.
%In this paper, we present our experimental results (summarized in \cite{Liu:13}) and discuss the theoretical foundation for the structures in Refs.~\onlinecite{Popovic07xing,Popovicxingpatent,Liu:13,Zhang:13}, including the low-loss unidirectional Bloch mode concept that is proposed in Refs.~\onlinecite{Popovic07xing,Popovicxingpatent}.
%In this paper, we present our experimental results on record loss in CMOS-compatible crossings %, demonstrate matching to theory, and present the low-loss unidirectional Bloch mode concept, suggested in %Refs.~\onlinecite{Popovic07xing,Popovicxingpatent},Ref.~\onlinecite{Popovic07xing}, which is the theoretical foundation for this work (and also applies to structures investigated in %Refs.~\onlinecite{Popovic07xing,Popovicxingpatent,Liu:13,Zhang:13,ShainlineOL13wigglermod}).
%Refs.~\onlinecite{Popovicxingpatent,Zhang:13,ShainlineOL13wigglermod}).
%Here, we provide more detail on our work in Ref.~\onlinecite{Liu:13}, and set a theoretical foundation for these structures.  Our results remain record low waveguide crossing loss in CMOS-compatible, photolithography patternable structures.%, and enable a path to lower loss designs through generalization to higher order designs.

% as the best waveguide loss patternable in CMOS photolithography (\cite{ePIXfab}, opsys, our 45nm work (100nm min feature size) ).  Near time of our submission, recent paper  \cite{Hochberg} has demonstrated 3x higher loss in similar designs following our proposal \cite{Popovic07xing}, and an improved design based on subwavelength gratings that achieves even lower loss \cite{Zhang:13}.  In this paper, we detail what we mean by the ``low-loss Bloch wave'' term we have assigned to this propagation mode, and explain in greater detail our recent results that demonstrated record waveguide crossing loss.

%In this paper, we experimentally demonstrate this concept showing near-theoretical performance with loss as low as $0.04\,$dB/crossing (1\%/crossing), and under $0.1$\,dB measured over a wide wavelength range exceeding  $100\,$nm.

%%%%%%%%%%%%%%%%%%%%%%%%%%%%%%%%%%%%%%%%%%%%%%%%
\begin{figure}[t]
\centering
\includegraphics[width=\linewidth]{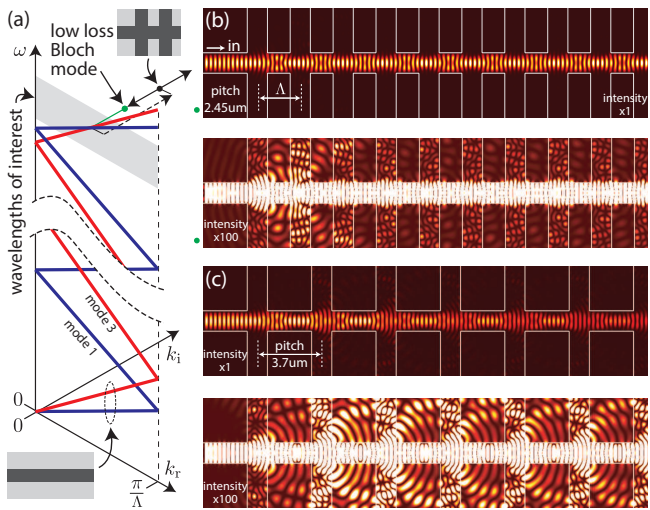}
\caption{Unidirectional low-loss Bloch modes: (a) illustration of complex-$k$ band structure of unperturbed waveguide and imaginary-$k$ splitting due to radiative crossings; (b) low-loss breathing field due to matching periodicities of the structure and of the breathing optical field [2D FDTD; bottom: 100x oversaturated intensity to show detail, green dot in (a)] \cite{Popovic07xing}; (c) high propagation loss with mismatched structure and breathing mode periodicities.\label{good}}
\vskip-10pt
\end{figure}

%\section{Design Approach}
The basic concept is similar to MMI-based \cite{Liu:04,Poon:06} synthesis of fields that focus across a waveguide crossing region \cite{Stuart:03} to prevent diffraction loss, extended to an open-system periodic array.  The periodicity gives rise to a unique, novel type of Bloch wave formed from two forward propagating modes of different transverse spatial order, which we can refer to as a unidirectional Bloch wave (Fig.~\ref{good}). %This kind of Bloch wave is formed of a superposition of the fundamental and third-order mode in a multimode waveguide, where a periodic perturbation (in this case the crossing waveguides) provides coupling between the two modes and leads to a bandgap.  
This Bloch wave differs from typical Bloch waves formed in photonic crystals by periodic coupling in that the half-wave periodicity of the PhC structure leads to bandgaps at the edges of the Brillouin zone, formed by an anticrossing between the dispersion curves of a forward and a backward guided mode.  In the present case, the periodicity leads to an anticrossing of a high order, between forward fundamental and third-order modes, which can be anywhere within the Brillouin zone [Fig.~\ref{good}(a)].  In this sense, the crossing array is more analogous to a long-period (fiber) grating than a photonic crystal.  
%A more unique aspect of 
The typical Brillouin-zone edge bandgap that is the key characteristic of photonic crystals will be primarily absent in these structures because they have nearly no periodicity-induced reflection.

The band structure is unusual in a second way.  The coupling at the waveguide-crossing points is not (only) a standard reactive, i.e. power-exchange, coupling that results from index perturbations,  but includes radiative interference coupling that scatters light from both the 1$^\mathrm{st}$ and 3$^\mathrm{rd}$ order transverse modes. %, leading to imaginary $k$-splitting.
%When this scattering interferes destructively, a low radiation loss eigenstate is established.  
%This means there is a propagation constant splitting that is not real, but imaginary (or, more generally complex when the standard index perturbation part is also accounted), and leads to one Bloch mode with a low imaginary part of the complex propagation constant, i.e. low propagation loss, and another with a high propagation loss.  There are analogous frequency eigenstates in resonators based on similar physics \cite{Dahlem10,Gentry13cleo}.
This means there is a propagation constant splitting that is not real, but imaginary (or, more generally complex when the standard index perturbation part is also accounted), and leads to one Bloch mode with a low imaginary part of the complex propagation constant (low loss), and another with a high loss.  There are analogous frequency eigenstates in resonators based on similar physics \cite{Dahlem10,Gentry13cleo}.

Physically, the low-loss Bloch mode corresponds to a superposition of modes 1 and 3 of the unperturbed waveguide with the right ratio of amplitudes to produce a field minimum (or null) at the scattering points. The high-loss mode in turn has constructive interference at these points. %the sidewalls at the scattering points.
%The unique insight is that the complex-propagation-constant eigenstates of the periodic structure uniquely select the field distribution with a high and low scattering loss.  Then, 
To realize a low-loss crossing array, the objective is to choose the waveguide cross-sectional dimensions, and crossing periodicity, that minimize the loss of the lower loss Bloch mode.  This insight also applies to the design of single crossings \cite{Liu:04,Popovicxingpatent,Hochberg}. % and should be the first step their design.
%, although thus far all designs except \cite{Popovic07xing} have treated the entire structure at once.

To arrive at a low-loss Bloch mode, our strategy is to maximize the imaginary splitting of the propagation constants for a \emph{given} scatterer geometry.  %One can think of scattering providing an imaginary part to the propagation constant that constitutes the radiation loss 
Radiation loss through scattering provides an imaginary part to the propagation constants [dashed arrow, Fig.~\ref{good}(a)], while radiative splitting brings back the low-loss mode down to as close to zero loss as the geometry permits via radiative cancellation 
[split arrows, Fig.~\ref{good}(a)].  As with real splitting, %for a given perturbation, 
the maximum splitting occurs at the phase matching condition.  
In our case, a periodic perturbation is required to match the difference in propagation constants of modes 1 and 3, 
%In the case of degenerate modes, a perturbation uniform along the waveguide suffices.  
%Since, in our structure, modes 1 and 3 differ in propagation constant, a periodic perturbation that matches this difference is required,
%, and leads to the optimal periodicity of the array.  Hence, to excite a low-loss breathing mode, the period of the crossing array is designed to first order to match the difference in (effective) propagation constants of the 1$^\mathrm{st}$- and 3$^\mathrm{rd}$-order eigenmodes of the waveguide, 
$\Lambda = 2\pi/(\beta_1-\beta_3)$.  Since $\beta_1$ and $\beta_3$ are eigenstates of the unperturbed waveguide without crossings, rigorous simulations that account for self-coupling perturbations \cite{PopovicCIFS} %of the crossings to the two modes lead to a small correction to the periodicity in the actual design.
in the crossings lead to a small correction to the periodicity in the actual design.

%The approach to design is then straightforward
%In a multimode waveguide without crossings, there are two Bloch waves modes of the form $$\mathbf e_{i}\,e^{j\Lambda z}= c_{i1}\cdot \mathbf  e_1\,e^{j\beta_1z}+c_{i2}\cdot \mathbf e_3\,e^{j\beta_2z},\ \ \  (i=a,b)$$ where $e_1$ and $e_3$ are the 1$^\mathrm{st}$- and 3$^\mathrm{rd}$-order eigenmodes of the straight multimode waveguide.  With crossings, scattering loss at the crossings acts as an imaginary $\delta\overline{\overline{\epsilon}}$ and radiatively couples these two modes following
%$$\mu_{ij}=\dfrac{\omega}{4}\displaystyle\int \mathbf e_i^*\cdot\delta\overline{\overline{\epsilon}}\cdot \mathbf e_j\ \text dV, \ \ \ (i,j=a,b)$$
%The resulting supermodes split along the imaginary axis in the complex frequency plane, and a low-loss (small imaginary eigen-frequency) mode that avoids the crossings is preserved, while the other one acquires high scattering loss and becomes suppressed.  %Launching the fundamental (Fig.~\ref{good}), in a periodicity matched structure, the low-loss Bloch wave survives as the wave spatially propagates along the crossing array.

%The Bloch wave picture is a helpful one is illustrated by considering the excitation of the multimode structure with only the fundamental mode of the waveguide in the wide section.  
If the fundamental mode is launched into the multimode guide [Fig.~\ref{good}(b)], the first few crossings strongly scatter until the low-loss wave is established.  The incident mode 1 field can be thought of as a superposition of the low-loss and high-loss Bloch waves with mode 1 components adding and mode 3 components canceling.  As the high-loss Bloch mode decays with a fast decay rate, only the low loss, breathing mode remains.  We can observe the fraction of mode 1 and 3 present in a cross-section where the low-loss wave has converged to a steady state.  If this ratio of modes 1 and 3 is excited at the start of the structure in Fig. 2b (as is done in Fig.~\ref{withtaper}), no power needs to be initially coupled to the high-loss Bloch wave, and thus no scattering will occur at the first few crossing points.

For efficient excitation of the low-loss Bloch wave from a single mode waveguide, we employ a symmetric, non-adiabatic taper.  It acts as a ``directional coupler'' between modes 1 and 3 of the multimode guide \cite{Liu:04,Popovic07xing,Popovicxingpatent} (by symmetry there is no coupling to asymmetric mode 2).
%Another approach is MMI excitation by overlap at an abrupt junction \cite{Poon:06}, accompanied by likely higher radiation loss.  The MMI design is equivalent to the right non-adiabatic taper design in the (non optimal) limit of zero taper length.
Another approach, MMI excitation at an abrupt junction \cite{Poon:06}, is equivalent to the non-adiabatic taper design in the (non-optimal) limit of zero taper length.

\begin{figure}
\begin{tabular}{p{.06\linewidth}p{.9\linewidth}}
(a)&
\begin{subfigure}[b]{\linewidth}
\raisebox{1em}{\raisebox{-\height}{\includegraphics[width=\linewidth]{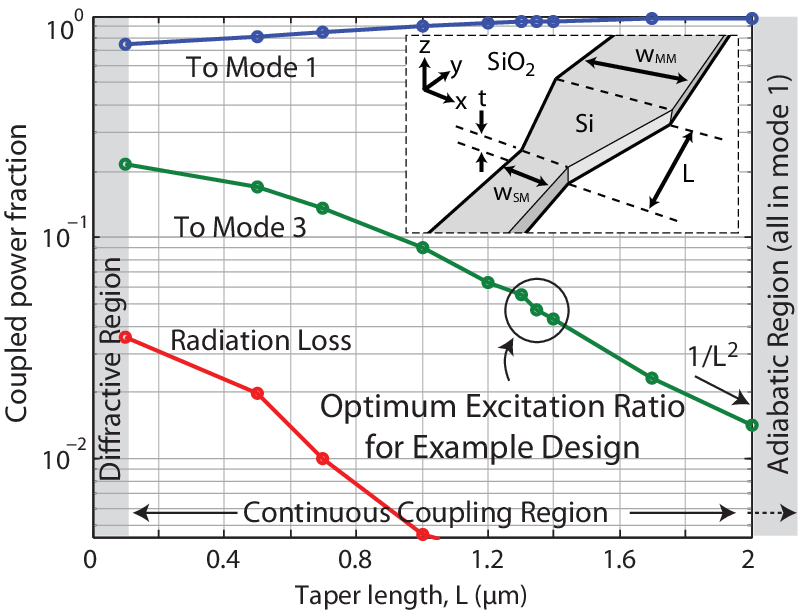}}}%\phantomsubcaption
\end{subfigure}\phantomsubcaption\label{taper}
\end{tabular}
\vskip3pt
\begin{tabular}{p{.06\linewidth}p{.9\linewidth}}
(b)&
\begin{subfigure}[b]{\linewidth}
\raisebox{1em}{\raisebox{-\height}{\includegraphics[width=\linewidth]{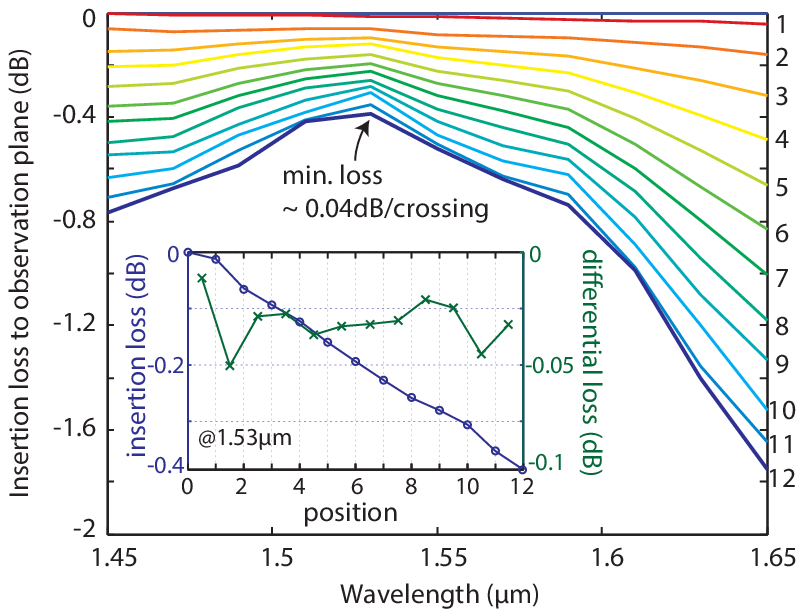}}}
\end{subfigure}\phantomsubcaption\label{fdtd}
\end{tabular}
\vskip-5pt
\begin{tabular}{p{.06\linewidth}p{.9\linewidth}}
(c)&
\begin{subfigure}[b]{\linewidth}
\raisebox{1em}{\raisebox{-\height}{\includegraphics[width=\linewidth]{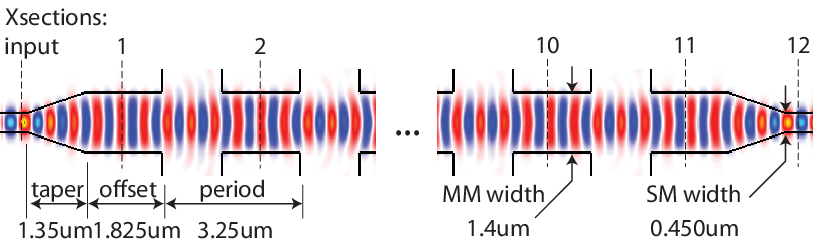}}}
\end{subfigure}\phantomsubcaption\label{field}
\end{tabular}

\label{fig:design}\vskip-10pt

\caption{Crossing design: \protect\subref{taper} short, non-adiabatic symmetric taper as a directional coupler between the $1^\mathrm{st}$- and $3^\mathrm{rd}$-order modes, allowing efficient excitation of a breathing Bloch mode of the crossing array; coupling and loss vs. length. \protect\subref{fdtd} Simulated transmission after 1-10 crossings, and through entire structure including tapers (3D FDTD). \subref{field} Field from a monochromatic 3D FDTD simulation at 1550\,nm.} %(dimensions shown differ slightly from the design used for fabricated devices).}
\vspace{-18pt}
\end{figure}

Devices for experimental demonstration were designed to be implemented in the $220$\,nm silicon device layer ($n=3.476$ at 1550\,nm wavelength) of a standard silicon photonics silicon-on-insulator (SOI) platform with a 2\,$\mu$m oxide undercladding, and oxide overcladding ($n=1.45$).  The width is chosen to sufficiently confine the first and third modes, without admitting a 5$^\mathrm{th}$ mode.  Dimensions are given in Fig.~\ref{field}, with the period obtained from a parameter sweep in 3D FDTD simulations to provide the minimum loss per crossing. % in the Bloch wave structure.

%A non-adiabatic taper can be designed to excite the optimum ratio of 1$^\mathrm{st}$ and 3$^\mathrm{rd}$ eigenmodes from a single-mode waveguide input.
%, to minimize loss.  This ensures that only the low-loss Bloch wave is excited.  
Figure~\ref{taper} shows 3D FDTD simulations of a linear taper design, showing coupling to mode 3, and corresponding radiative loss.  
For an optimum mode 3 coupling ratio of just under 5\%, a 1.35\,$\mu$m taper suffices, and the radiation loss is negligible ($<$0.4\%) so we didn't consider more complex taper shapes. % here.  %In general, tapers that gradually rather than abruptly couple modes 1 and 3, i.e. distribute coupling along the taper can minimize coupling to radiation modes.  
The taper in Fig.~\ref{taper} is broadband, with 4.75\% coupling at 1550\,nm, a 0.3\% variation over 1450--1550\,nm, increasing to 6\% at 1650nm.

Figure \ref{fdtd} shows the simulated transmission spectra of a complete waveguide crossing array with 10 crossings and input/output tapers to 450\,nm-wide, single mode waveguides.  The insertion loss vs. wavelength is computed after 1--10 crossings, with observation planes labeled in Figs.~\ref{fdtd} (right) and \ref{field}, and through the entire structure including tapers, using 3D FDTD.  
%Theoretical transmission is $\sim 0.4\,$dB (average 0.04\,dB/crossing) at the design wavelength, including the two tapers, and $\sim$0.033\,dB average per added crossing.  
Peak theoretical transmission is $\sim 0.4\,$dB for 10 crossings and two tapers, and $\sim$0.033\,dB average per added crossing (Fig. \ref{fdtd} inset).  
Figure \ref{field} shows the transverse electric field distribution in the structure designed to excite 
%exciting the low-loss Bloch wave with a monochromatic 1550\,nm source in the fundamental mode of the input guide. 
the low-loss Bloch wave from a single mode % is excited with the fundamental mode of the 
input guide at $1550\,$nm.
The input taper must be placed at a particular offset from the first crossing in order for the first ``focal point'' of the field envelope to coincide with the first crossing waveguide.  %This is found from the relative phases of the two modes exiting the taper, and the beat length necessary to arrive at the relative phase of the low-loss Bloch wave's focal point.

%%%%%%%%%%%%%%%%%%%%%%%%%%%%%%%%%%%%%%%%%%%%%%%%
%\vspace{-6pt}
%\section{Experimental Results}
%Figures \ref{om} and \ref{sem} show the optical microscope and SEM images of a waveguide crossing array. 

\begin{figure}
\begin{tabular}{p{.06\linewidth}p{.9\linewidth}}
(a)&
\begin{subfigure}[t]{\linewidth}
\raisebox{1em}{\raisebox{-\height}{\includegraphics[width=\linewidth]{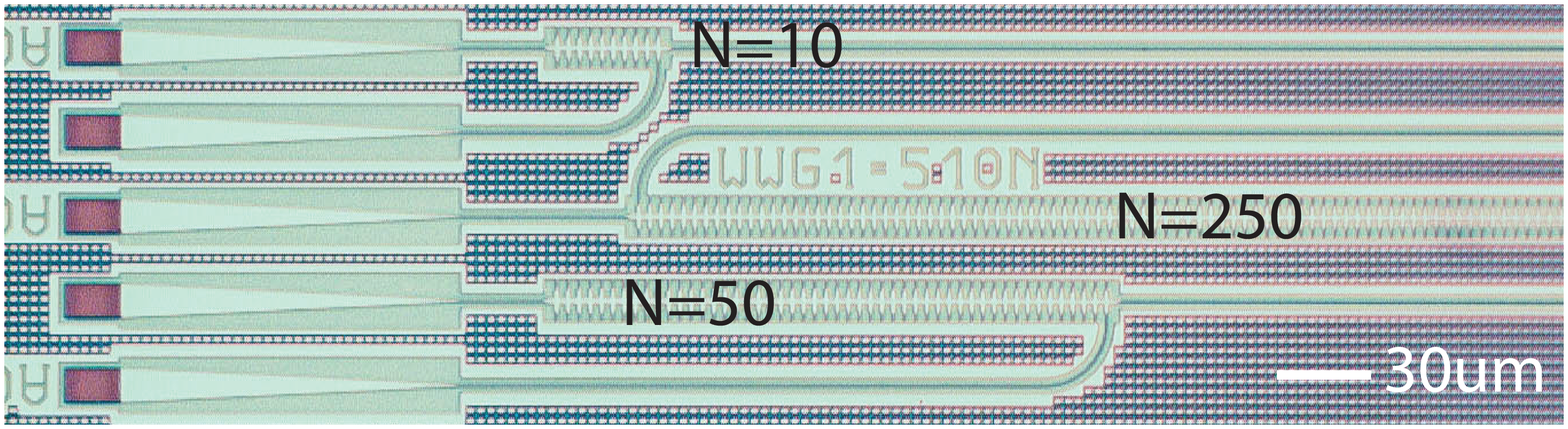}}}
\end{subfigure}\phantomsubcaption\label{micrograph}
\end{tabular}
\vskip3pt
\begin{tabular}{p{.06\linewidth}p{.9\linewidth}}
(b)&
\begin{subfigure}[t]{\linewidth}
\raisebox{1em}{\raisebox{-\height}{\includegraphics[width=\linewidth]{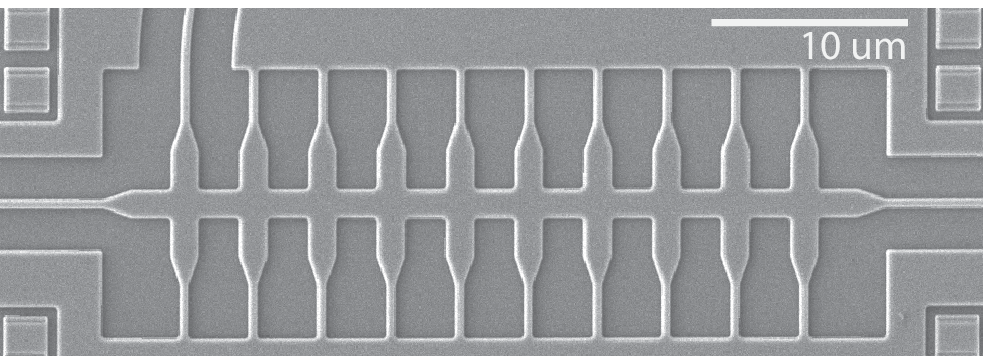}}}
\end{subfigure}\phantomsubcaption\label{SEM1}
\end{tabular}
\vskip3pt
\begin{tabular}{p{.06\linewidth}p{.9\linewidth}}
(c)&
\begin{subfigure}[t]{\linewidth}
\raisebox{1em}{\raisebox{-\height}{\includegraphics[width=\linewidth]{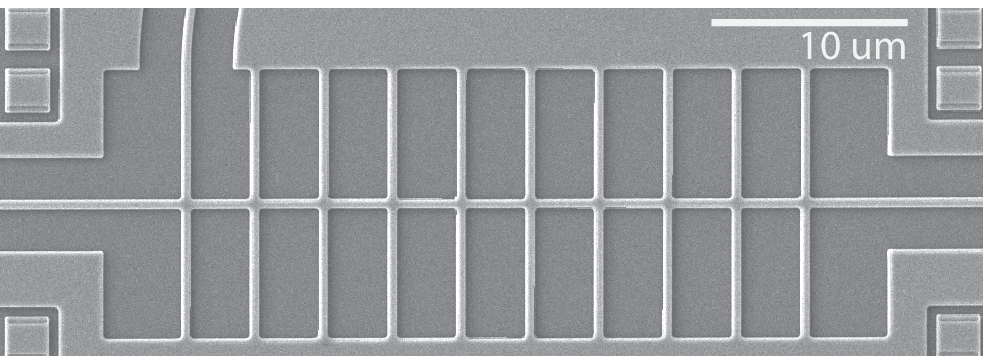}}}
\end{subfigure}\phantomsubcaption\label{SEM2}
\end{tabular}

\label{fig:image}%\vskip-5pt
\caption{\subref{micrograph} Optical micrograph of on-chip crossing arrays fabricated through IMEC/ePIXfab \cite{ePIXfab} with 10, 250 and 50 periods. \subref{SEM1} SEM of a 10-period low-loss Bloch crossing array. \subref{SEM2} SEM of a 10-period normal crossing array.}
\vskip-15pt
\end{figure}

Arrays with $N = 10, 50, 150$ and $250$ crossings were realized.  Fig.~\ref{micrograph} shows a typical set of on-chip devices, fabricated through the ePIXfab multiproject wafer service \cite{ePIXfab}.  %Target design dimensions are given in Fig.~\ref{field}.  
To pre-compensate for bias in the lithography, the width of the multimode waveguide section was varied in layout from $1450\,$nm to $1510\,$nm, with other dimensions remaining at design target (Fig.~\ref{field}).  Figures ~\ref{SEM1} and \ref{SEM2} show scanning electron micrographs (SEMs) of a Bloch-wave 10-crossing array and a ``normal'' 10-crossing array based on $450$\,nm-wide single mode waveguides.%, respectively.

\begin{figure}
\begin{tabular}{p{.06\linewidth}p{.9\linewidth}}
(a)&
\begin{subfigure}[t]{\linewidth}
\raisebox{1em}{\raisebox{-\height}{\includegraphics[width=\linewidth,trim=0 0 0 0cm,clip]{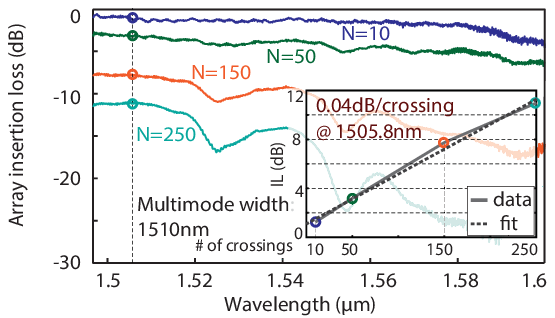}}}
\end{subfigure}\phantomsubcaption\label{spec_all}
\end{tabular}

\begin{tabular}{p{.06\linewidth}p{.9\linewidth}}
(b)&
\begin{subfigure}[t]{\linewidth}
\raisebox{1em}{\raisebox{-\height}{\includegraphics[width=\linewidth,trim=0 0 0 0.05cm,clip]{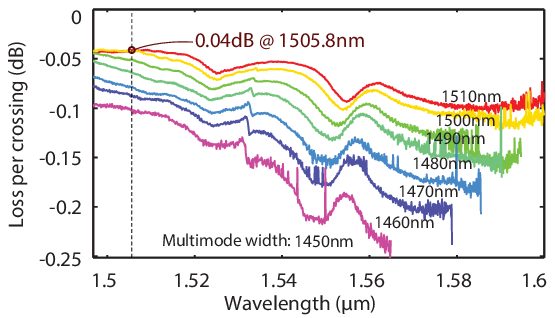}}}
\end{subfigure}\phantomsubcaption\label{loss_per_X}
\end{tabular}

\label{fig:experiment}\vskip-5pt
\caption{%\protect\subref{om} Optical micrograph of three waveguides with $50$, $250$ and $10$ crossings. \protect\subref{sem} SEM detail of crossing arrays. 
\protect\subref{spec_all} Spectral insertion loss of Bloch arrays with $10$, $50$, $150$ and $250$ crossings (1510\,nm width); \protect\subref{loss_per_X} insertion loss/crossing vs. wavelength of crossing arrays with waveguide widths from $1450\,$nm to $1510\,$nm.}
\vskip-15pt
\end{figure}

Figure~\ref{spec_all} shows measured total insertion loss %experimentally measured in these devices, 
normalized to %a plain single mode waveguide of equal total length.  Thus, input and output grating coupler responses are common-moded
%normalized 
%out of the spectra.  The data represents 
%represent the insertion loss of the array, 
include losses from the non-adiabatic tapers and the crossings, relative to a plain single mode waveguide. %Here, insertion loss means the additional cost of replacing a single mode waveguide with the crossing array; hence, in principle, if there is substantial propagation loss due to sidewall roughness, by this definition it is possible for the insertion loss to be slightly negative, if the multimode waveguide has lower sidewall roughness loss, and the crossings are near lossless.
For each waveguide width variant, the slope of insertion loss versus number of crossings gives the loss per crossing (not including the tapers) at each wavelength (linear fits, see Fig.~\ref{spec_all} inset, give Fig.~\ref{loss_per_X}).  A device with waveguide width $1510\,$nm shows an average loss per crossing of $0.04\,$dB at $\lambda=1505.8$\,nm.  The flatter loss slope of the last 100 crossings (Fig.~\ref{spec_all}, inset) shows loss of {0.032\,dB/crossing with $\sim\!\pm0.005$\,dB uncertainty} due to coupling (equivalent to $\sim\!100\,$dB/cm distributed loss). %SEMs of the 1510\,nm wide structure (Fig.~\ref{sem}) confirm that its measured width is near the targeted 1400\,nm.
%The other major source of difference from theoretical design is error in the device layer thickness.  
The demonstrated loss %of 0.04\,dB/crossing 
matches theory closely, but the minimum loss wavelength is shifted down by about 25\,nm.  This is consistent with an error in device layer thickness/width.  For small shifts, the minimum insertion loss is not substantially changed.%, but the wavelength at which it occurs changes.

%\begin{figure}
%\centering
%
%\begin{subfigure}[t]{.5\linewidth}
%\includegraphics[width=\linewidth]{4a.eps}\vskip-5pt
%\caption{}\label{spec_all}
%\end{subfigure}\hskip-.1pt
%\begin{subfigure}[t]{.5\linewidth}
%\includegraphics[width=\linewidth]{4b.eps}\vskip-5pt
%\caption{}\label{loss_per_X}
%\end{subfigure}
%
%\begin{subfigure}[b]{.5\linewidth}
%\includegraphics[width=\linewidth]{4c.eps}\vskip-5pt
%\caption{}\label{loss_vs_num}
%\end{subfigure}\hskip-.1pt
%\begin{subfigure}[b]{.5\linewidth}
%\includegraphics[width=\linewidth]{4d.eps}\vskip-5pt
%\caption{}\label{TX}
%\end{subfigure}
%
%\label{fig:experiment}%\vskip-5pt
%\caption{
%\protect\subref{spec_all} Spectral insertion loss of arrays with $10$, $50$, $150$ and $250$ crossings (waveguide width varies from $1450\,$nm to $1510\,$nm). 
%\protect\subref{loss_per_X} Insertion loss/crossing vs. wavelength of crossing arrays with different waveguide width.  \protect\subref{loss_vs_num} Insertion loss vs. number of crossings  in designed crossing array of waveguide width $1510\,$nm compared with an array of $10$ normal crossings, at $\lambda=1.5058\,\mu$m where the designed crossing achieves the lowest loss of $0.04\,$dB/crossing, and at $\lambda=1.5421\mu$m where the normal crossing array has the lowest loss. \protect\subref{TX} Transmission and crosstalk spectra of designed and normal crossing arrays with 10 periods.}
%\vspace{-6pt}
%\end{figure}

\begin{figure}

\begin{tabular}{p{.06\linewidth}p{.9\linewidth}}
(a)&
\begin{subfigure}[b]{\linewidth}
\raisebox{1em}{\raisebox{-\height}{\includegraphics[width=\linewidth,trim=0 0 0 0cm,clip]{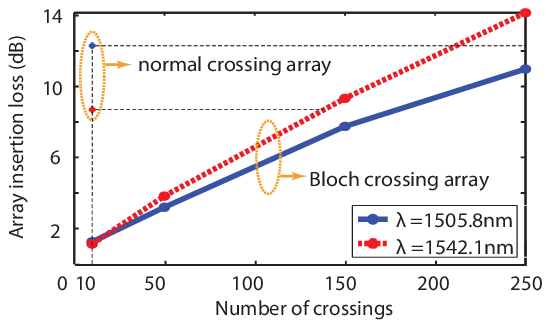}}}
\end{subfigure}\phantomsubcaption{}\label{loss_vs_num}
\end{tabular}

\begin{tabular}{p{.06\linewidth}p{.9\linewidth}}
(b)&
\begin{subfigure}[b]{\linewidth}
\raisebox{1em}{\raisebox{-\height}{\includegraphics[width=\linewidth,trim=0 0 0 0.05cm,clip]{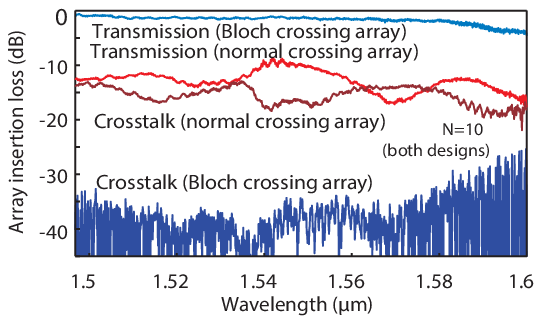}}}
\end{subfigure}\phantomsubcaption{}\label{TX}
\end{tabular}
%\phantomsubcaption\label{spec_all}
%\phantomsubcaption\label{loss_per_X}
%\phantomsubcaption\label{loss_vs_num}
%\phantomsubcaption\label{TX}}

\label{fig:compare}\vskip-5pt
\caption{\protect\subref{loss_vs_num} Insertion loss vs. number of crossings in Bloch crossing array (width $1510\,$nm) and 
%an array of $10$ normal crossings, 
a 10-crossing normal array, 
at $\lambda=1505.8\,$nm where the Bloch crossing has the lowest loss and at $\lambda=1542.1\,$nm where the normal crossings have the lowest loss. \protect\subref{TX} Transmission and crosstalk spectra of the Bloch and normal crossing arrays with 10 periods.}
\vskip-10pt
\end{figure}

%Comparing this optimal design with a plain, single-mode $10$-crossing array, 
At the $250$-crossing Bloch array's optimal wavelength (Fig.~\ref{loss_vs_num}), it has lower insertion loss than a $10$-crossing single-mode array.  %At the single-mode crossing's best performing wavelength ($\lambda=1542.1$\,nm, green), the multimode crossing array still out performs significantly.
Figure~\ref{TX} shows the transmission and crosstalk of a $1510\,$nm-wide 10-crossing Bloch-wave design and a 10-crossing single-mode array.  Apart from $13\,$dB higher transmission, approaching sub-1\% loss per crossing, the Bloch-wave crossing array also suppresses crosstalk by at least 35\,dB (limited by measurement noise), 20\,dB more than a normal crossing array.  This is consistent with the theoretical prediction \cite{Popovic07xing}. %in the initial proposal of these structures \cite{Popovic07xing}.

%%%%%%%%%%%%%%%%%%%%%%%%%%%%%%%%%%%%%%%%%%%%%%%%
%\section{Conclusion}
The experimental demonstration of CMOS compatible waveguide crossing arrays with ultra-low losses, matching theory, may be enabling and impact photonic network-on-chip architecture. % \cite{Batten,Joshi,Bergman14}.  
These structures offer capabilities for electrically/thermally active, suspended and optomechanical photonic structures with minimized scattering loss.  More generally, this concept extends to higher order, as well as transversely asymmetric structures \cite{Popovicxingpatent,Liu:13_2},  as recently applied in modulators \cite{ShainlineOL13wigglermod}.

%\noindent {\bf Acknowledgment:} 
\noindent This work was supported by NSF award ECCS-1128709.

\end{document}